\documentclass{vgtc}                          

\graphicspath{{figures/}{pictures/}{images/}{./}} 

\usepackage{times}                     

\usepackage{tabu}                      
\usepackage{booktabs}                  
\usepackage{lipsum}                    
\usepackage{mwe}                       
\usepackage{amsmath}
\usepackage{amsfonts}

\usepackage{mathptmx}                  

\onlineid{0}

\vgtccategory{Research}

\vgtcinsertpkg

\title{Tracking the Spatiotemporal Spread of the Ohio Overdose Epidemic with Topological Data Analysis}

\author{Nicholas Bermingham \thanks{e-mail: bermingham.11@osu.edu}\\ %
        \scriptsize The Ohio State University %
\and David White\thanks{e-mail: whiteda@denison.edu}\\ %
     \scriptsize Denison University %
\and Nathan Willey \thanks{e-mail: willey.106@osu.edu}\\ %
     \parbox{1.4in}{\scriptsize \centering The Ohio State University}}

\teaser{
  \centering
    \includegraphics[width=.8\linewidth,]{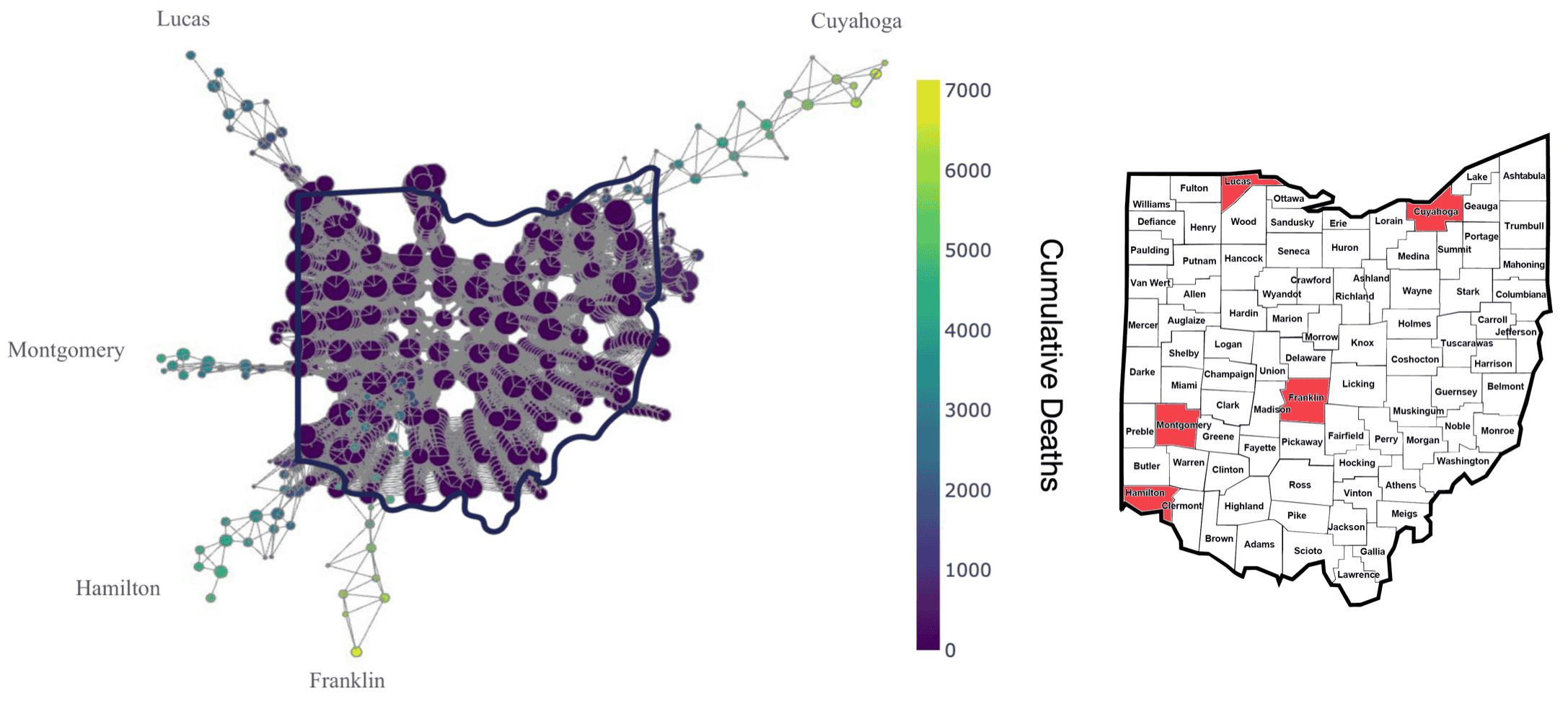}
    \caption{Mapper graph created from county-level statistics of cumulative drug overdose deaths next to a map of Ohio with counties that form flares in the Mapper graph highlighted. The geometry of the Mapper graph corresponds to the geography of Ohio and flares in the graph highlight death counts that are substantially different from those of the surrounding counties.}
    \label{fig:AnnotatedMapper}
}

\abstract{
    In recent years, techniques from Topological Data Analysis (TDA) have proven effective at capturing spatial features of multidimensional data. However, applying TDA to spatiotemporal data remains relatively underexplored. In this work, we extend previous studies of disease spread by using the Mapper algorithm to analyze the Ohio drug overdose epidemic from 2007 to 2024. We introduce a novel method for constructing covers in Mapper graphs of spatiotemporal data that respects geographic structure and highlights the time-dependent variables. Finally, we generate a Mapper visualization of regional demographics to examine how these factors relate to overdose deaths. Our approach effectively reveals temporal trends, overdose hotspots, and time‑lagged patterns in relation to both geography and community demographics.
}


\begin{document}


\firstsection{Introduction}

\maketitle

In this paper we investigate the effectiveness of applying the topological data analysis (TDA) tool known as Mapper to data relating to the spatiotemporal aspects of the Ohio overdose epidemic from January $2007$ to September $2024$. We examine what insights into the Ohio overdose epidemic can be drawn from the topological and geometric features of Mapper graphs, such as connected components, flares and holes.

Drug overdose remains a leading cause of preventable death in the United States. Between 1999 and 2022, annual overdose deaths rose more than sixfold, and in 2022 alone, over 107,000 Americans died from drug overdoses. The impact is not uniform: overdose mortality varies widely across time, geography, and demographic groups. Some of the hardest-hit areas, including counties in Ohio, have experienced multiple waves of the epidemic, each marked by distinct spatial and temporal patterns. As synthetic opioids make the drug supply increasingly volatile, it is important to understand where and when risk is increasing, and how it's spreading from one county to another. Modeling the spatiotemporal spread of overdose deaths can help public health agencies identify emerging hotspots early, allocate limited resources more effectively, and implement tailored interventions before crises peak. In a policy environment of shrinking federal support, tools that can capture the evolving geography of the epidemic are critical to preventing future deaths.

The spread of overdose deaths does not follow simple trajectories: it is shaped by changes in drug supply, local economic distress, demographic shifts, and public health infrastructure. Traditional approaches to modeling these patterns—including spatial regression and generalized linear mixed models—require strong assumptions about linearity, locality, and parametric form. High-dimensional models with large numbers of county-specific coefficients are often difficult to interpret and may obscure broader structural features. This landscape calls for flexible, non-parametric tools—like those from TDA—that can reveal large-scale structure in complex, high-dimensional data without relying on restrictive modeling assumptions.

The Mapper algorithm is a tool, introduced in \cite{singh2007topological} and widely used in TDA, to construct graphs that encapsulate the broad topological features of a given dataset. Mapper constructs a simplified representation of a high-dimensional dataset by clustering nearby observations and connecting them into a graph that reflects both global and local structure. This allows one to more easily visualize data and draw both qualitative and quantitative insights.

We apply the Mapper algorithm to county-level geographic, demographic, and drug overdose death counts in Ohio across a seventeen-year period with the goal of understanding the spatiotemporal spread of overdose deaths. 
Our methodology is informed by \cite{chen2021topological}, a similar analysis for Covid-19 spread using Mapper.
To the best of our knowledge, drug overdose data have not yet been analyzed with TDA and so one of our goals is to determine its utility for both this specific application and more general spatiotemporal analyses. 
Additionally, we extend the experiments of \cite{chen2021topological} in a few ways.
First, we explore whether there is more insight to gain from utilizing population-normalized death counts to better represent data from rural communities. Second, we use a novel method for selecting Mapper hyperparameters to best capture the evolution of spatiotemporal data. 
Third, going beyond spatiotemporal analysis, we investigate whether similar insights can be gained from building Mapper graphs from county-level demographic data and overdose data to create a ``demographic landscape" of Ohio.
We use visual properties of the resulting graphs in each setting to identify patterns of spread, population-normalized burden, and demographic correlates.


Our findings confirm and extend known patterns in Ohio’s overdose epidemic. We identify the emergence of rural hotspots near the Ohio River and the increasing burden in counties that experienced high unemployment. We also observe a time-dependent trajectory in which overdose mortality spreads outward from urban centers into surrounding regions. These results align with prior work using conventional modeling frameworks (see \hyperref[subsec:related-works]{Section 1.2}). Notably, we find evidence of a lagged association between economic stressors and subsequent overdose mortality, as well as visual confirmation of unique regional trajectories (e.g., in Scioto County) consistent with known historical events and public health reports.

By producing intuitive and interactive visualizations that highlight where and how the overdose crisis has unfolded, our approach has implications for future public health planning. Spatiotemporal analytics in Rhode Island helped motivate the state’s implementation of geographically targeted interventions, with overdose prevention resources deployed at the neighborhood level \cite{marshall2022preventing}. In that context, visual evidence of spatial spread prompted decision-makers to shift from reactive policies to proactive containment. We hope the ideas in this work can serve a similar function in Ohio and other states, both by providing a new lens on the overdose crisis and by demonstrating the potential of TDA methods in public health research.

\subsection{Outline of the Paper}

In \hyperref[section:methods]{Section 2}, we first outline the Mapper algorithm and the features that allow us to customize it for specific analysis of spatiotemporal data.
Next we discuss how we have sourced and organized our data on spatiotemporal spread of overdose deaths and county-level demographics in Ohio. 
We then detail the specific way we have implemented the Mapper algorithm to analyze these datasets, discuss how we have implemented a novel choice of hyperparameters to create more intuitive and interpretable Mapper graphs and note how our implementation differs from the work done in \cite{chen2021topological}.

In \hyperref[section:results]{Section 3}, we display our Mapper graphs, analyze the factors contributing to the emergence of topological and geometric features in the graphs and identify the counties whose data contributes to the emergence of these features.

Finally, in \hyperref[section:conclusions]{Section 4} we draw conclusions regarding what our investigations tell us about the Ohio overdose epidemic, how our different methodologies could be used to study similar temporal datasets, and future directions.

\subsection{Related Works}
\label{subsec:related-works}

This paper represents the first use of topological data analysis to visualize drug overdose data at the time of writing.
Previous works have modeled the temporal autocorrelation of drug overdose in Ohio using time-series analysis methods, such as \cite{bci} which created one model for all of Ohio, determining how the monthly number of deaths $D_t$ depends on its own past.
More generally, \cite{rosenblum2020rapidly} applied a generalized linear mixed model (GLMM) to opioid overdose death counts $D_{c,t}$ in each (county, month) pair in Ohio.
While this method is general enough to account for both spatial and temporal dependence, since $D_{c,t}$ can be a function of any other $D_{c',t'}$, it faces challenges due to the large number of parameters involved.
In \cite{rosenblum2020rapidly}, the potential for one county to affect its geographical neighbors is not explored and linear relationships between each $(c,t)$ and $(c,t-h)$ are assumed.
Verifying linearity assumptions for these numerous relationships can be cumbersome, requiring looking at one scatterplot for each coefficient in the model.
These challenges motivated our current use of TDA techniques, which allow for the exploration of high-dimensional spatiotemporal data without relying on strict parametric assumptions or individual coefficient estimation.


Several strands of research have also addressed the spatial dynamics of the drug overdose epidemic in Ohio. Andrew Curtis and members of the Begun Center for Violence Prevention have fit spatial models for drug overdose data in the Cleveland area \cite{curtis2025using, mcmaster2025drug}, at the census block level. 
This kind of technique can produce heat maps and cartographic maps showing which areas are most at risk of overdose spikes, e.g., showing movement of the epidemic into African American neighborhoods by comparing heat maps in one year with the next year. However, the statistical models do not include the time dimension, and we are unaware of how these models can be used for forecasting future hotspots. Additionally, Adam Eck and his students at Oberlin College use machine learning models (e.g., random forests, gradient boosting, individual decision trees, SVMs, neural networks) to predict county-level overdose death hotspots \cite{eck}, with the explicit aim of helping guide public policy and resource allocation.

In addition, it is possible to approach spatial and spatiotemporal autocorrelation using a Bayesian framework. Kline, Hepler, and their students have employed Bayesian statistical models to estimate spatial autocorrelation in opioid overdose deaths across Ohio counties, providing insight into geographic clustering and county-level risk factors \cite{hepler2019latent, kline2021estimating, kline2021multivariate}. These papers fit generalized spatial factor models, and look at the relationship between treatments for substance use disorder and drug overdose deaths, in each county. Their algorithm produces spatial weights for each county, interpreted as the degree of unmeasured heterogeneity across counties, causing statistically significant differences that the model cannot explain. This work was generalized to add a temporal dimension by Ji \cite{ji2019joint}. Others have fit similar models in the Cincinnati area \cite{li2019suspected, choi2022spatial}, at the census block level.

There has also been work done at the national level and in other states, e.g., \cite{stewart2017geospatial}. The most advanced appears to be Rhode Island, where academic researchers have teamed up with the state health department to develop the PROVIDENT system \cite{marshall2022preventing}. This system uses both machine learning algorithms and statistical models (e.g., spatiotemporal Gaussian processes) to predict future hotspots at the census block level using SUDORS data (explained in \cite{sudors-ohio}).
The state health department uses these predictions to optimize their deployment of overdose prevention resources at the neighborhood level.

Beyond statistical models, it is also possible to model spatiotemporal spread using Hawkes processes and other methods from dynamical systems.
The middle author used these models to determine the spatiotemporal spread of protests in the USA \cite{rodriguez2023analysis} and in Ukraine \cite{bahid2024statistical}.



Numerous previous papers have applied TDA to other epidemics including the spatiotemporal spread of Covid-19 \cite{chen2021topological, hickok2022analysis, ault2022comparison}, Zika \cite{lo2018modeling, soliman2020ensemble, rudkin2023spatial}, influenza \cite{costatopological}, and other contagious diseases \cite{ taylor2015topological}. Although the mechanisms of spread of overdoses from county to county are very different from the spread of a virus from person to person, the success of these previous applications of TDA inspired our current analysis, especially \cite{chen2021topological}.


\section{Methods}
\label{section:methods}

\subsection{Mapper Algorithm}

The Mapper algorithm is a versatile TDA tool developed by Singh, M\'emoli and Carlsson \cite{singh2007topological} for qualitative analysis and visualization of high-dimensional datasets in a way that preserves topological features.
The Mapper algorithm has been used in a variety of fields, such as economics where it has been used to detect inter-dependencies of factors involved in firm financial ratios \cite{dlotko2024financial}, environmental science where it has been used to study factors contributing to harmful algal blooms and many other areas discussed in Section 3.1 of \cite{MapperAlgorithmReview}. 
Chen and Volic's paper \cite{chen2021topological} showed that Mapper could visually capture the emergence of Covid-19 hotspots and provide insights about the spatial and temporal links between them through analysis of the flares and holes in the Mapper graphs.

\begin{figure}[tb]
    \centering
    \includegraphics[width=1\linewidth]{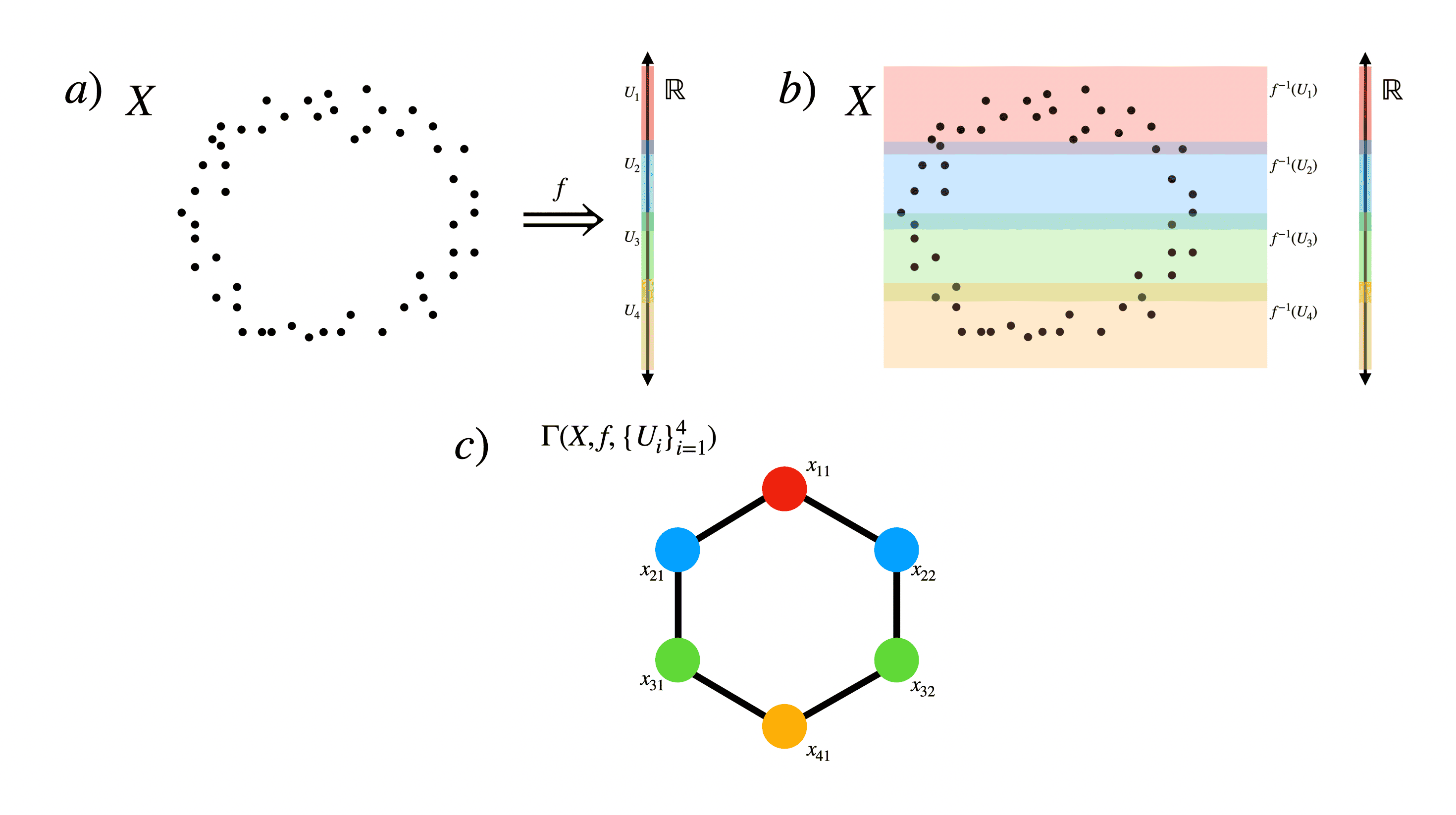}
    \caption{a) A dataset $X\subset \mathbb{R}^2$ being mapped onto $\mathbb{R}$ via the projection function $f(x,y) = y$ and a covering of $\text{im}(f)$ by 4 open sets $U_1,U_2,U_3,U_4$. b) The preimages $f^{-1}(U_i)$ overlaid over the original dataset $X$. c) The resulting Mapper graph whose nodes $x_{ij}$ come from applying a clustering algorithm to $X\cap f^{-1}(U_i)$ to find the related data points, and whose edges come from the nodes whose intersection $x_{ij}\cap x_{kl} \neq \emptyset$.}
    \label{fig:MapperInfographic}
\end{figure}

The Mapper algorithm works by taking a potentially high-dimensional dataset $X\subset \mathbb{R}^N$ and projecting this dataset onto a lower-dimensional space via a chosen filter function or lens, $f:X\to \mathbb{R}^d$. The algorithm then finds a finite cover of the range of the function via a methodology specified by the user. A clustering algorithm, again specified by the user, is then used on the preimage, $f^{-1}(U_i)$, of each $U_i$ in the cover to create a set of nodes $\{x_{ij}\}_{j=1}^{n_i}$. These clusters then appear in the Mapper graph as vertices representing parts of our dataset that the clustering algorithm deemed to be close together. Edges between two nodes $x_{ij}$ and $x_{kl}$ are drawn in the Mapper graph whenever their intersection $x_{ij}\cap x_{kl} \subset X$ is nonempty.  This graph is then visualized as an interactive plot in either 2 or 3 dimensions using standard parts of the {\tt plotly} package.  A simple example of this procedure which shows how the Mapper graph captures topological features of a given dataset is presented in \cref{fig:MapperInfographic}.

The main strength of Mapper is its versatility and customizability for a researcher's needs,  evident in the number of choices in the Mapper pipeline that are left to the user. However this also leads to Mapper's biggest weakness as a data analysis tool: its sensitivity to the choice of filter function, cover parameters, and clustering algorithm \cite{singh2007topological, MapperAlgorithmReview}.  Certain variants of Mapper such as Ball Mapper and F-Mapper have been developed to address these sensitivity issues but these improvements come at the cost of interpretability and computational complexity \cite{MapperAlgorithmReview}. As interpretability is our primary concern in this paper, we opt to use the standard Mapper algorithm and have tried to be principled in our choice of these parameters to minimize sensitivity issues which will be discussed further in \hyperref[subsec:Implementation]{Section 2.3}.


\subsection{Ohio Overdose and Demographic Data}

Our dataset of drug-induced deaths in Ohio comes from DataOhio, which tracks death records from the Ohio Department of Health's (ODH) Bureau of Vital Statistics \cite{dataohio}. These data are reported monthly and our analysis covers the period from January 2007 to September 2024. It is important to note that the actual number of overdose deaths might not match the number in our dataset, e.g., because drug overdose is sometimes unreported as a cause of death on death certificates \cite{buchanich2018effect}. The ODH and the Centers for Disease Control try to correct for this, but missing data remains a potential concern. The population data used for the Mapper graphs is yearly and comes from the Census 10-year estimates \cite{census-pop}. 

For the spatiotemporal results, the data points used to create the Mapper graphs are $4$-dimensional and of the following form:
\[\big(\text{month}, \text{latitude}, \text{longitude}, \text{(normalized) cumulative deaths}\big).\]
Each data point represents one month's data for a given county whose center is the reported latitude and longitude.

In order to visualize the spread of the Ohio drug epidemic throughout the demographic regions of Ohio, we selected key demographic indicators that we used to form the following 8-dimensional data points:
\begin{align*}
    \big(\text{year}, &\text{population}, \text{\% poverty}, \text{median-income}, \text{\% unemployed},\\
    &\text{\% white},\text{cumulative deaths},\text{normalized cumulative deaths}\big).
\end{align*}
Our choice of demographic features to consider was informed by other works \cite{kline2021estimating, rosenblum2020rapidly} which determined these features to be relevant to understanding issues pertaining to drug use in Ohio.
Both the percent of the population in poverty and the median income of a county were sourced from the Census SAIPE program \cite{saipe}. Data on the percent of the population that is white and total population count come from the US Census American Community Survey \cite{acs}. Data from both sources are tracked yearly at the county level.
Unemployment data came from the Ohio Department of Job and Family Services through the Local Area Unemployment Statistics program \cite{laus}. Unemployment data is tracked monthly at the county level.
For visualization purposes, these features are aggregated to be yearly estimates.
We used data from the years 2009-2023 inclusive. 
This change from the monthly data used for the spatiotemporal results is done for the sake of visual clarity in the final Mapper graphs.
Since we used publicly available data, with no identifying features, no IRB was required.

We remark that the work done in \cite{chen2021topological} calls for renormalizing the data columns to ensure that no aspect of the data is weighted disproportionately.
In our chosen implementation of Mapper, the cubical cover divides the range of each column into $n$ intervals of equal size, regardless of prior normalization, thus the resulting Mapper graph is unchanged, and we do not need column-normalization.
In contrast to traditional statistical methods, the Mapper algorithm is relatively robust to data points with identical values in some of the coordinates.
This is important given that our data points use features gathered on different time-scales; for example yearly population data with monthly death counts with static latitude and longitude.
As we will discuss later, the predictable and static structure of the latitude, longitude, and time coordinates can help to determine hyperparameters for our Mapper graphs and provide a data-driven approach to the visualizations.

\subsection{Implementation of Mapper for the Ohio Overdose Epidemic}
\label{subsec:Implementation}

Our implementation of Mapper is done in Python through the {\tt mapper} functions available in the {\tt giotto-tda} package \cite{tauzin2020giottotda}. 
Our filter functions are always projections onto a certain subset of the data that is relevant to each investigation we undertake. 
We use projections, as opposed to other classical choices like density and eccentricity estimators \cite{singh2007topological}, both to reproduce the findings of \cite{chen2021topological} and to aid in interpreting the results. By using projection functions and not applying any other transformations to our data, we know that data within a cluster and adjacent clusters must have similar values in the features that we project onto.
To cover the image of our filter function we choose a cubical cover which requires the user to specify a number of intervals $n$, and a percentage overlap $p$. The cover is then constructed by dividing the range of the function in each dimension $i = 1,...,d$ into $n$ intervals of equal size $\mathcal{I}_i=\{I_{ij}\}_{j=1}^n$ such that the percentage of $I_{ij}$ which overlaps with $I_{i(j+1)}$ is $p$ for all $j<n$. The resulting cover consists of high-dimensional rectangles of the form $ \prod_{i=1}^d I_{ij_i}$ where $j_i \in \{1,...,n\}$ for all $i$. The values $n$ and $p$ are hyperparameters which typically require experimentation in order to find the optimal values to use.
We use DBSCAN clustering which is the default clustering algorithm recommended in {\tt giotto-tda}. The advantages of DBSCAN in comparison to more naive clustering algorithms are discussed in \cite{schubert2017dbscan}. Specific comparison of DBSCAN to other methods for building Mapper graphs can be found in \cite{MapperAlgorithmReview} which finds DBSCAN to be ``ideal for datasets with distinct but continuous structures" which is the case for our datasets.
Finally, to produce visuals of the Mapper graphs we choose to embed them into $\mathbb{R}^3$ rather than $\mathbb{R}^2$ in order to improve the clarity of our figures. Embeddings into $\mathbb{R}^2$ often produced cluttered visuals with distinct connected components overlapping in ways that made them appear connected. While our 3D plots also have points that are occluded by the embedding, they provide more information especially when viewed through the interactive versions provided in the \hyperref[sec:supplemental_materials]{supplementary materials}.

\begin{figure}[tb]
    \centering
    \includegraphics[width=0.9\linewidth]{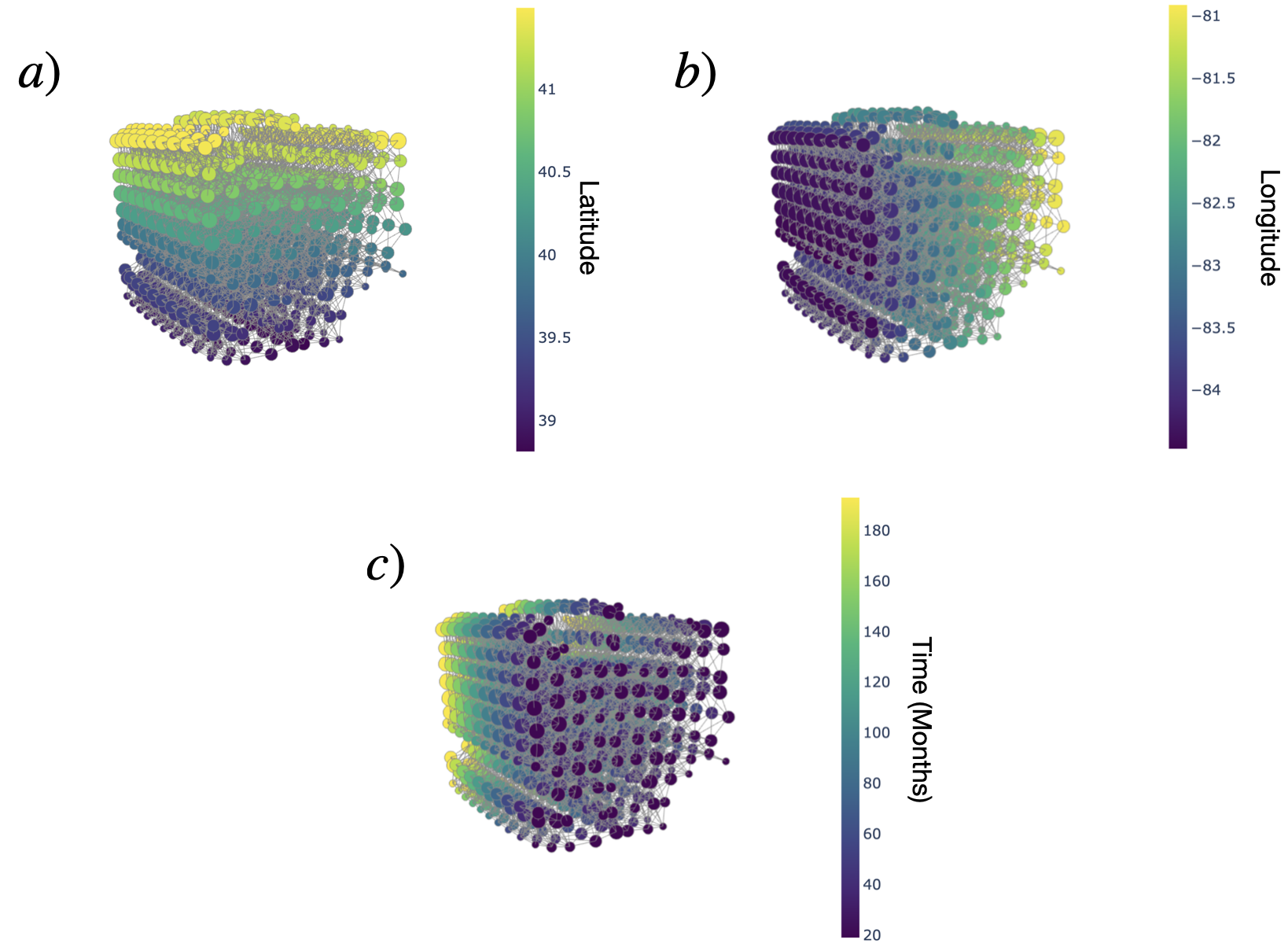}
    \caption{The Mapper graph formed by projecting onto just our spatiotemporal data with our informed choice of hyperparameters. Nodes are colored by the cluster averages of a) latitude, b) longitude, c) month since January 2007.}
    \label{fig:GeographicPlots}
\end{figure}

Our first goal is to reproduce the findings of \cite{chen2021topological} which showed that flares in certain Mapper graphs could identify hotspots in an epidemic. Following \cite{chen2021topological} we use the 4-dimensional dataset containing cumulative death counts in a county, as detailed in the previous section. We use cumulative deaths as opposed to death count in a given month because their almost-continuous nature aligns better with the geometric aspects of the Mapper graph. Notably, the steady increase of this parameter over time leads to the formation of a flare in the Mapper graph, called a `chain of nodes' in \cite{chen2021topological}. 
For our cover of this spatiotemporal data, we implement a novel and informed choice of hyperparameters. The main observation informing our choice is that without the death information, the data used for our spatiotemporal plots simply captures the geography of Ohio, represented by the latitude and longitude of each county of Ohio, staying constant over time. As Ohio is a contiguous land mass and our time is varying continuously, the most representative plot of this data would be one totally connected component that varies in two spatial dimensions and persists unchanged in a time dimension. Thus we choose the hyperparameters $n$ and $p$ so that the Mapper graph formed by projecting onto only spatiotemporal data represents this connectivity correctly with the minimal overlap required. In doing so, we obtain a Mapper graph (embedded in $\mathbb{R}^3$) whose vertical slices represent the adjacency of Ohio's counties and whose third axis only captures the flow of time. Following this approach led us to using $10$ intervals with $50\%$ overlap. The resulting Mapper graph obtained is depicted in \cref{fig:GeographicPlots}. 

This informed choice of hyperparameters aids in reproducibility of our methods. 
While choosing parameters resulting in a fully connected Mapper graph of the spatiotemporal variables is nontrivial and results from experimentation, ensuring that these choices visually capture the geographic information of an area being studied should always be possible and will result in similar findings. 
This choice also aids in interpreting the resulting Mapper graphs because when we add overdose information into our projection our cover guarantee that any topological or geometric features that appear (such as holes or flares) are caused only by the death data and not noise in the geographic location of county centers. 

Our next investigation is into whether this methodology extends well to population-normalized statistics. 
For this, we use the same choice of hyperparameters on our 4-dimensional dataset containing cumulative deaths normalized by county population, instead of regular cumulative deaths. 
We study whether areas of relatively high deaths are similarly identifiable as flares in the Mapper graph and how the Ohio overdose epidemic played out in low population regions that are more difficult to observe when looking at raw cumulative death counts.

Finally, we investigate how to use Mapper to study the relationship between the demographic profiles within the counties of Ohio and the rates of overdose deaths. 
We first build a demographic landscape of Ohio using the demographic dataset detailed in the previous section. 
Our filter function is initially projecting on the six factors of year, population, \% poverty, median-income, \% unemployed and \% white, and we adjust the hyperparameters of our cover to create the most useful and informative demographic plot we can. 
Unlike the spatiotemporal setting, these choices are more subjective and our choice was broadly informed by our knowledge of the demographics of Ohio consisting of a handful of metropolitan areas contrasting against a very large number of regional and rural communities. We found that using $9$ intervals with $45\%$ overlap gave graphs that represented this kind of demography appropriately and allowed for some interesting insights to be drawn.
Having constructed this demographic landscape we consider two different ways to incorporate overdose death data into our analysis. The first is to simply color the demographic landscape according to (normalized) cumulative deaths and the second is to reconstruct the Mapper graphs incorporating (normalized) cumulative deaths into the factors we projected onto while keeping our hyperparameters the same. We then evaluate the differences in the two methodologies and discuss insights we can gain from these plots into the Ohio overdose epidemic.

Our methodologies here serve two principal aims: first, to show how Mapper can be used in an experimental fashion to see the effect of a variable on the Mapper graph given a ``control" graph, and second, to investigate the nature of the spatiotemporal spread of overdose deaths in Ohio by analyzing these changes.

\section{Results}
\label{section:results}

In this section, we present the Mapper graphs created by the methods discussed in the above section. We use these graphs to analyze trends in overdose deaths in Ohio. We divide our results into three main parts: spatiotemporal analysis based on cumulative death counts, spatiotemporal analysis based on population-normalized counts, and an analysis of how demographic landscapes relate to overdose deaths. HTML versions of the 3D embeddings used to construct these figures, in which you can change the viewing angle and magnification, can be found in the \hyperref[sec:supplemental_materials]{supplementary materials}. 
These may assist in viewing certain features of the figures presented here.

\subsection{Spatiotemporal Visualization of Overdose Deaths}

\begin{figure}[tb]
    \centering
    \includegraphics[width=\linewidth]{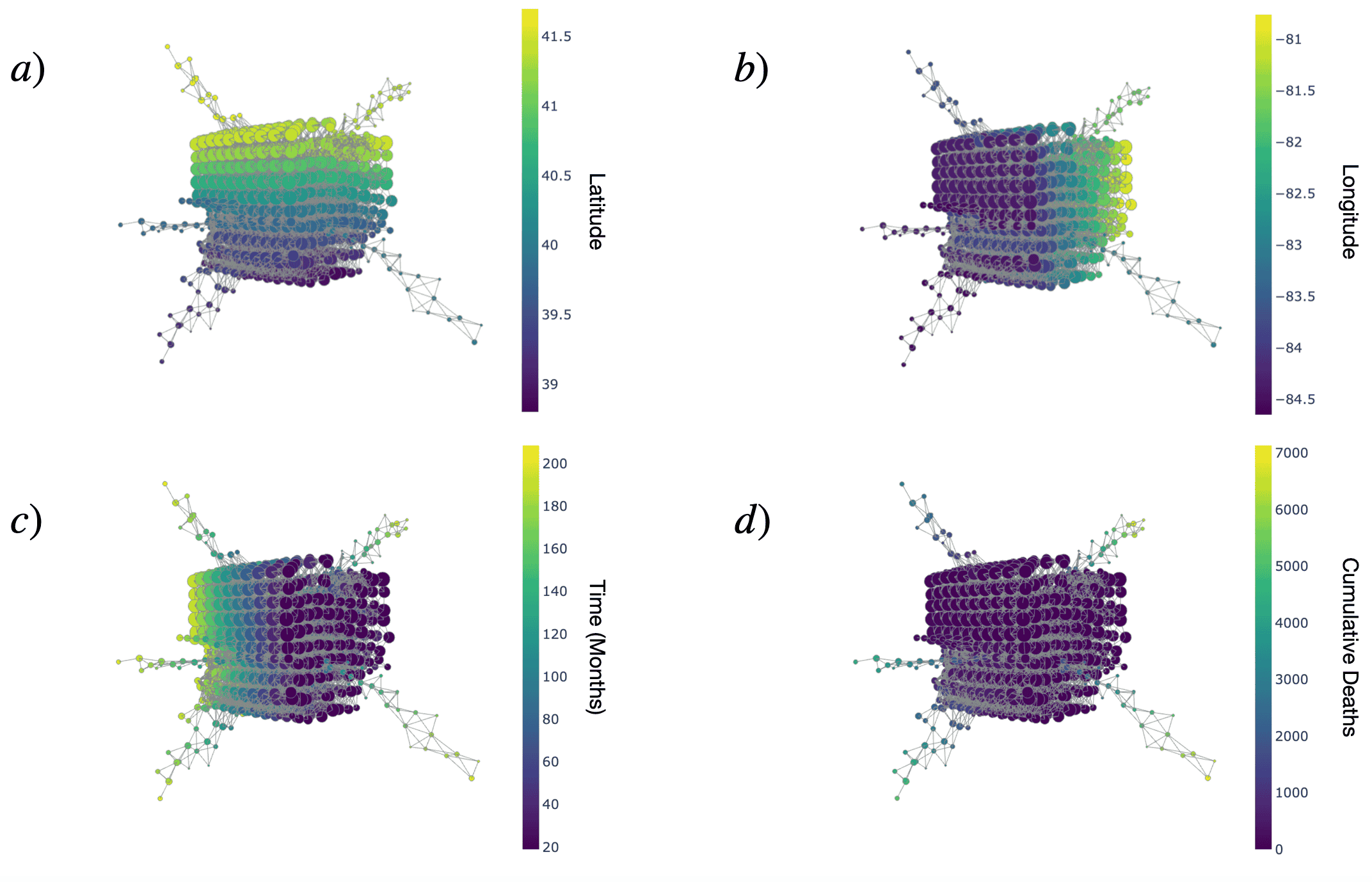}
    \caption{The Mapper graph formed by the spatiotemporal data along with cumulative deaths; the visually protruding flares show large increases in overdoses with time. 
    Nodes are colored by the cluster averages of a) latitude, b) longitude, c) month since January 2007, d) cumulative number of deaths due to drug overdoses. }
    \label{fig:CumulativeDeathPlots}
\end{figure}

We begin by using the $4$-dimensional cumulative death dataset with the filter function set to the identity on $\mathbb{R}^4$.
The Mapper graph created from this dataset is shown in \cref{fig:CumulativeDeathPlots} using the previously detailed hyperparameters. 

It is clear that the broad structural features of this graph are inherited from the graph constructed without death information (\cref{fig:GeographicPlots}) as the graphs still predominantly capture the geographic position of counties of Ohio persisting over time. We will refer to this piece of the Mapper graph corresponding to Ohio over time as the ``main trunk.''

The main geometric difference created by incorporating cumulative overdose death information into the projection filter function is that counties with high populations (those containing major urban areas of Ohio) create visible flares or branches extending from the main trunk (See \cref{fig:CumulativeDeathPlots}). By analyzing color changes in the flares in \cref{fig:CumulativeDeathPlots} c) and d), we can tell that the flares go away from the main trunk with increasing time and death values, capturing how these counties experienced a constantly increasing cumulative death count over time that substantially differed from the death counts in surrounding counties.
This successfully reproduces the findings of \cite{chen2021topological} that major population centers were the main driver of flares in the Mapper graphs constructed from their cumulative counts. A labeled diagram demonstrating the correspondence between the geography of Ohio and the geometry of the graph is presented in \cref{fig:AnnotatedMapper}. The labels were determined by analyzing the nodes that contribute to these branches off of the main trunk, which showed that each branch was comprised solely from the data of one county. The counties that branched off were: Franklin (containing the city of Columbus), Lucas (containing Toledo), Hamilton (containing Cincinnati), Montgomery (containing Dayton), and Cuyahoga (containing Cleveland).

Analysis of these emergent topological features of the Mapper graph gives us information about how our data is evolving locally in space and time. The origin of these flares determines the geographic location of the county in which a spike in death data occurs. A flare whose nodes contain only information from one county and does not reconnect with the main trunk indicates a strong trend in death count relative to the surrounding area. A flare with more connected nodes containing data from multiple counties indicates a spatial cluster of counties whose death totals are evolving similarly in time.

Using these ideas, we can analyze the specific flares in our Mapper graph to find county-level trends in the overdose epidemic. From the coloring in \cref{fig:CumulativeDeathPlots}~c) we can see, for example, that Franklin County starts to spike around 2015 (month 100).
As it never reconnects with the main trunk and the nodes only contain data from Franklin County, this indicates that the death count stays high relative to the surrounding counties during subsequent months.
Although not all are visible from the angle in this figure, many of the other main population centers follow a similar timeline to Franklin County in overdose death spikes. However, in the case of the Cincinnati area we observe a hole at the base of the main flare formed by nodes containing data from the nearby Butler County connecting to this branch from the main trunk. This corresponds to the cumulative death count in this county temporarily catching up to where Hamilton County was in its earlier death count, before the cumulative death count of Hamilton County substantially increased. These cross-county features agree with previously observed trends of overdose deaths occurring at high rates in the cluster of counties around the Cincinnati area, while staying more concentrated around Columbus. This also demonstrates the Mapper algorithm's ability to illustrate spread in time across many counties together.

Franklin County illustrates one limitation of Mapper graphs. Due to its central location in Ohio, when plotted it is visually difficult to see the exact month at which the flare peels off of the main trunk. However, a careful study of the interactive plots in the \hyperref[sec:supplemental_materials]{supplementary materials} does show this information.

This analysis agrees with the kinds of results obtained in \cite{chen2021topological} confirming the utility of Mapper graphs in analyzing data relating to spatiotemporal spread of epidemics.
The trends in spatiotemporal data most amenable to analysis via Mapper graphs are those that are relevant to the geometry and topology of the dataset, notably the connected components, flares, and holes created by a cumulative variable.
For the cumulative death count, as expected, we see that flares correspond to large population centers which have large death counts.
In order to further explore the capabilities of Mapper in analyzing this kind of spatiotemporal data we next look at rates of overdose to reduce the impact of high population areas on these visualizations.

\subsection{Spatiotemporal Visualization of Population-Normalized Overdose Deaths}

\begin{figure}[tb]
    \centering
    \includegraphics[width=\linewidth]{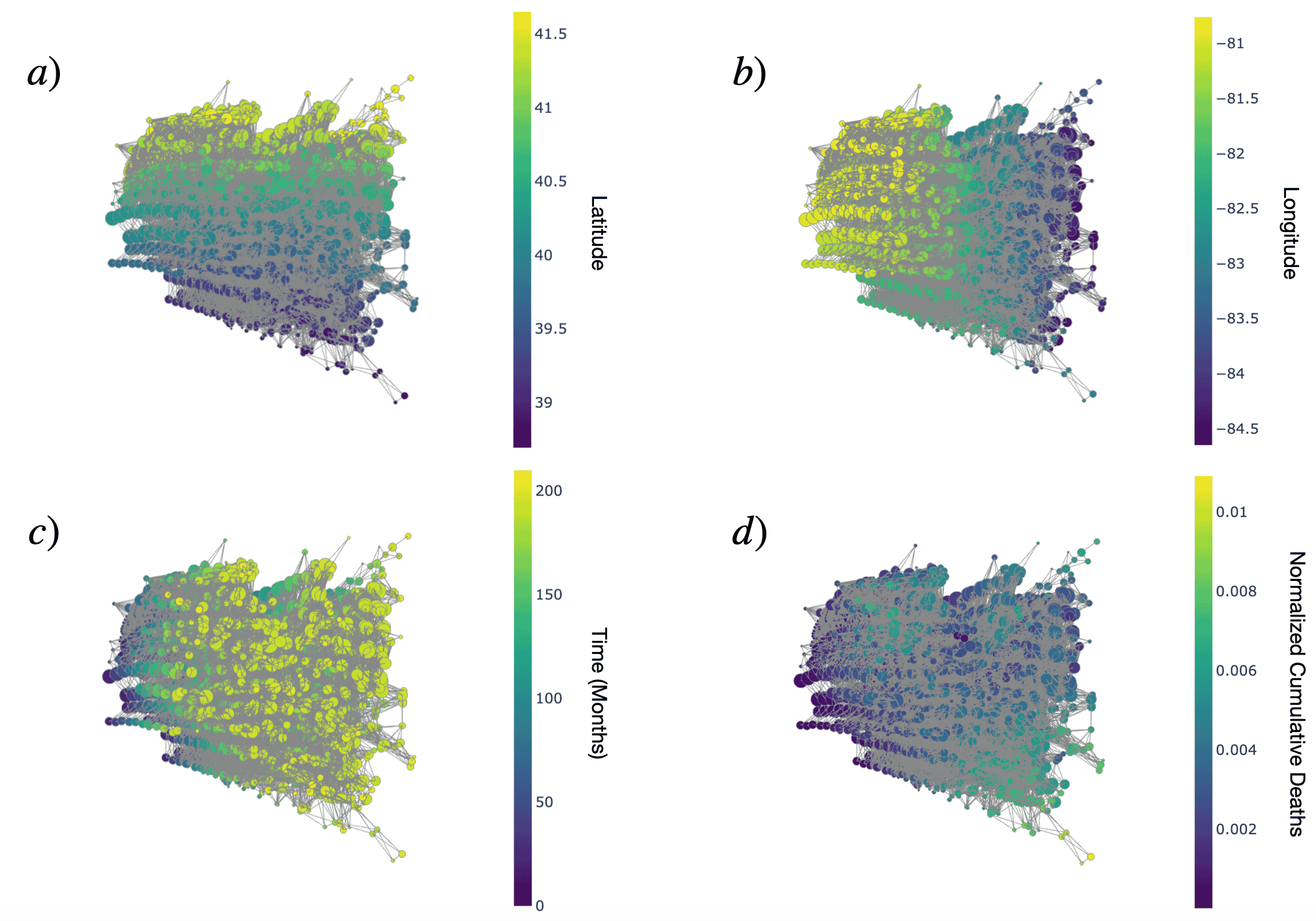}
    \caption{The Mapper graph formed by our spatiotemporal data with cumulative deaths normalized by the population of the county where the deaths occurred. West to east runs right to left here.\\
    The coloration of the nodes is by the cluster averages of a) latitude, b) longitude, c) Month since January 2007, and d) Cumulative number of deaths due to drug overdoses normalized by county population.}
    \label{fig:NormalizedCumulativeDeathPlots}
\end{figure}

We now consider the Mapper graph constructed from the 4-dimensional dataset containing cumulative death counts normalized by county population with a filter function set to the identity map. This plot is presented in \cref{fig:NormalizedCumulativeDeathPlots}.

We note that the images in \cref{fig:NormalizedCumulativeDeathPlots} are viewed from the opposite perspective to those in \cref{fig:CumulativeDeathPlots}. This is done in order to highlight the flares formed by higher normalized cumulative deaths which occur predominantly towards the end of the time period analyzed. As a result the latitude of the plot appears reversed from the traditional east-west perspective highlighted in \cref{fig:AnnotatedMapper}.

\begin{figure}[tb]
    \centering
    \includegraphics[width=\linewidth]{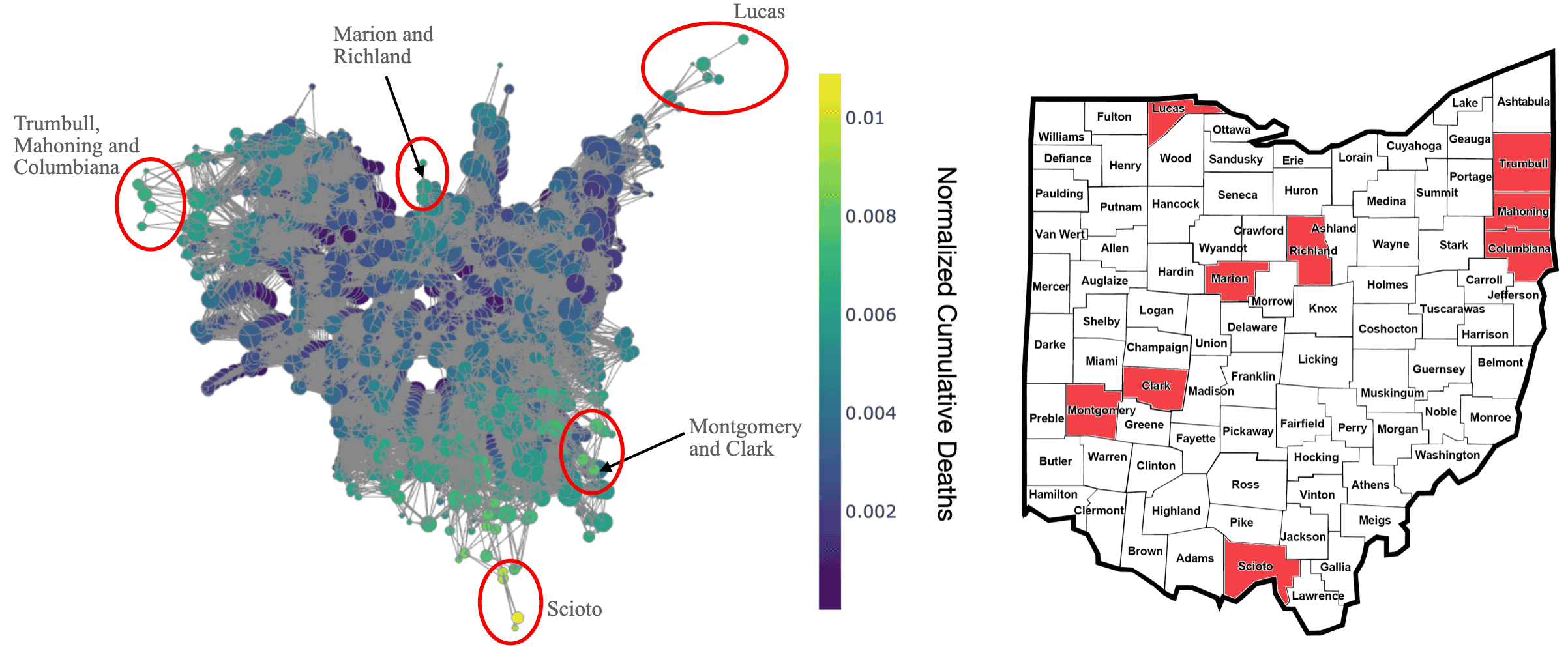}
    \caption{A front view of the Mapper graph in \cref{fig:NormalizedCumulativeDeathPlots}. Flares are labeled by the counties whose data is contained within the clusters which are highlighted on the map to the right of this plot.}
    \label{fig:labelednormalizeddeaths}
\end{figure}

The immediate visual takeaway is how much more chaotic \cref{fig:NormalizedCumulativeDeathPlots} appears to be compared to \cref{fig:CumulativeDeathPlots}, suggesting that the normalization process creates less distinction between counties than the unnormalized data. 
This makes sense because population is a major distinguishing feature between counties. However, we do clearly see some flares forming in the data, most prominently in southern Ohio near the Ohio River which forms the border between Ohio and Kentucky, and Ohio and West Virginia.
A labeled diagram of this Mapper graph is presented in \cref{fig:labelednormalizeddeaths} which points to Scioto County as having the highest value of normalized cumulative deaths. There are also a number of smaller peaks in areas around cities like Marion and Richland counties north of Columbus, Mahoning, Trumbull and Columbiana just East of Cleveland, Montgomery and Clark, containing and near Dayton, and Lucas containing Toledo.

This finding aligns with the county-by-county analysis of \cite{rosenblum2020rapidly} as well as the middle author's investigation of overdose death rates using SUDORS data \cite{sudors-ohio}. These works identified counties in Appalachia, in the Dayton-Cincinnati corridor, and south and east of Cleveland, as areas with higher than average overdose death rates. For that reason, those same counties have been a focal point for harm reduction efforts in Ohio, such as the HEALing Communities project \cite{el2020introduction}.

Another important feature is that these population-normalized flares occur later in time than the large city-center flares of the previous section, generally after 150 months whereas the flares in cities start between month 80 and 120.
The timing of these smaller peaks occurring in low-density areas relative to the peaks from cities shown in \cref{fig:CumulativeDeathPlots} may suggest a form of delayed spread. One interpretation of this is that the drug epidemic starts to spike in cities and later on flows into less populous areas of Ohio where the death toll is more significant relative to the county population.

This shows that we can extend the findings of \cite{chen2021topological} to normalized death counts and obtain new insights into our data by analyzing the same kinds of topological and geometric features.
Additionally, by creating and comparing Mapper graphs built from both raw death count and population-normalized count we get further insights into the nature of the Ohio overdose epidemic, its spatiotemporal spread, and how this intersects with population dynamics.

\subsection{Mapper Visualization of Demographic Data and Overdose Deaths}

Finally, we analyze how demographic shifts over time correspond with the Ohio overdose epidemic. These results are exploratory in nature and expand on the utility of the Mapper algorithm as a way to investigate temporal data in relation to a variety of features.

\subsubsection{Visualizing the Demographic Landscape}

\begin{figure}[tb]
    \centering
    \includegraphics[width=1\linewidth]{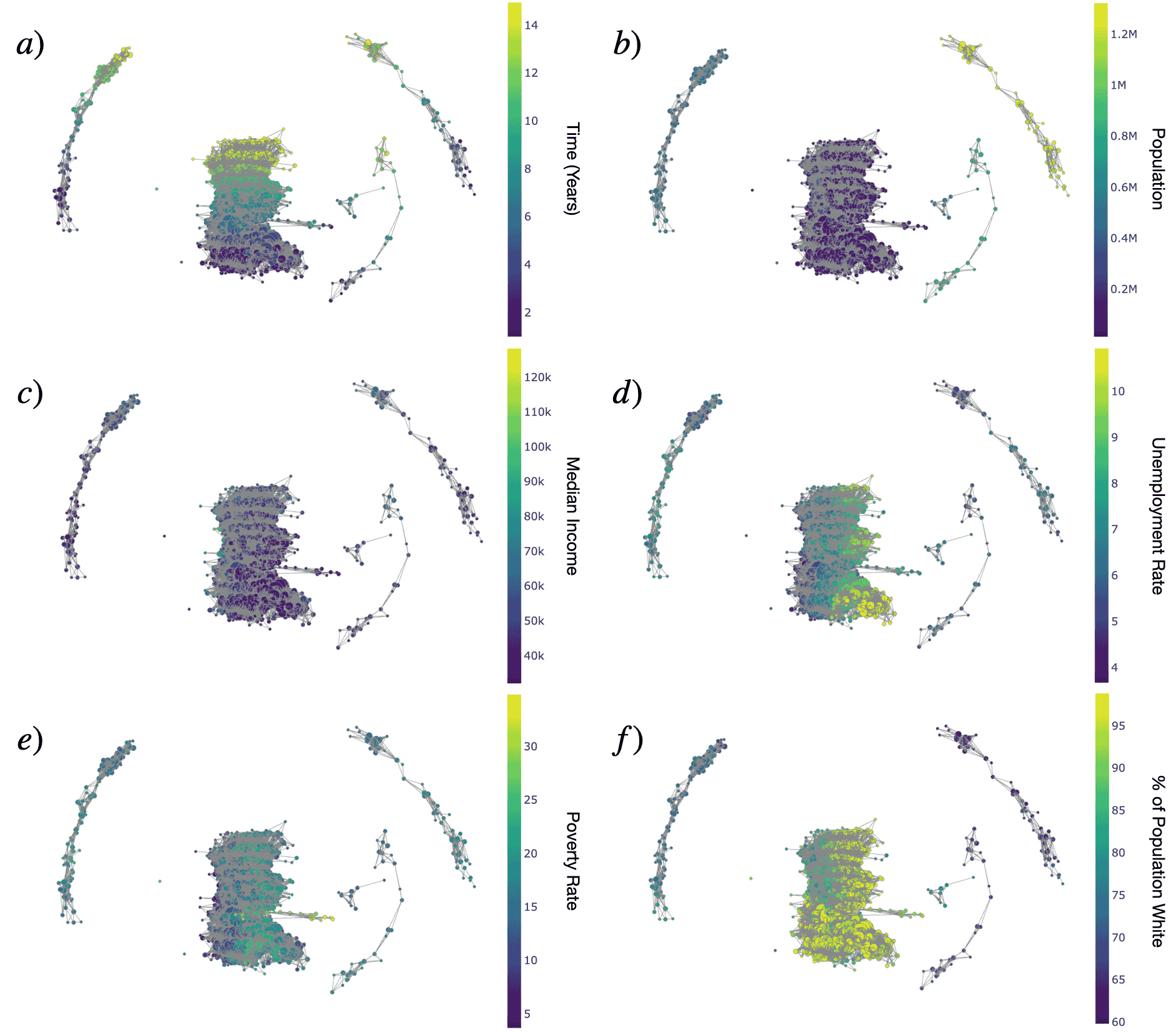}
    \caption{Mapper graphs of the demographic profile of Ohio colored according to a) Time in years, b) Population, c) Median Income, d) Unemployment rate, e) Poverty rate, f) Percentage of the county that is white.}
    \label{fig:demographic-mapper}
\end{figure}

We begin by constructing a Mapper graph from the demographic information described in \cref{section:methods}, a snapshot of which is depicted in \cref{fig:demographic-mapper}, with nodes colored by each parameter projected onto by the filter function. Recall that close and connected nodes in this visualization represent counties with similar demographic profiles at similar times.

\begin{figure}[tb]
    \centering
    \includegraphics[width=\linewidth]{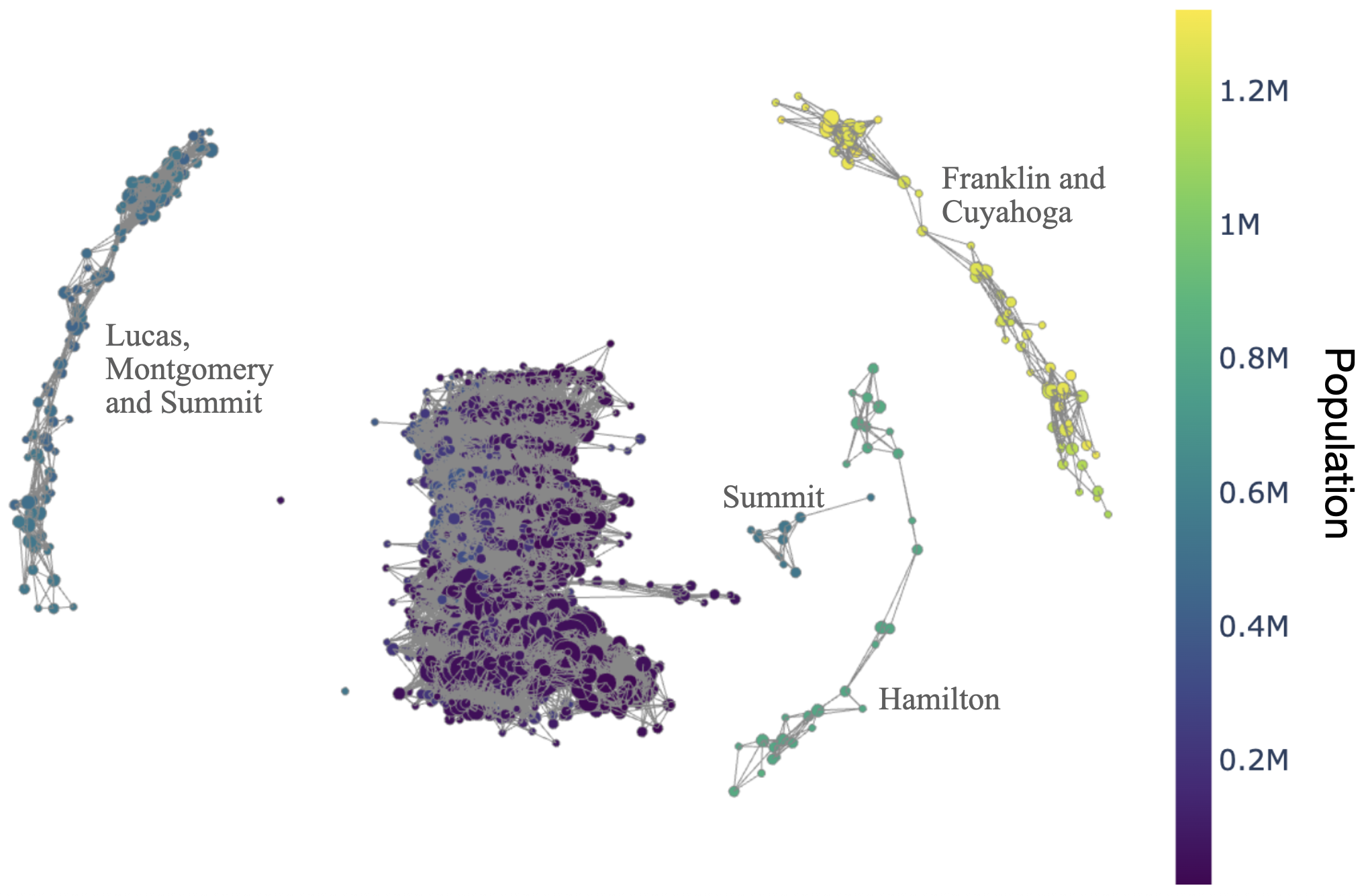}
    \caption{The demographic plot in \cref{fig:demographic-mapper} b) labeled to identify the counties whose data contributes to each of the smaller connected components.}
    \label{fig:labelleddemographicplot}
\end{figure}

One of the immediate observations that we can draw from these figures is that the connected components seem to be clearly delineated by population. This is observable by noting the uniformity of color in the connected components of \cref{fig:demographic-mapper} b) and makes sense as population is the statistic with the greatest range and delineation between counties of all those considered here.
By investigating the clusters that constitute these connected components, we can determine that one of these components is comprised of data from Franklin and Cuyahoga, the most populous counties in Ohio containing the cities of Columbus and Cleveland respectively. Another connected component is composed solely of data from Hamilton County containing the city of Cincinnati next to one smaller connected component comprised of data from Summit County containing the city of Akron. The final skinny connected component is comprised of data from  Summit, Lucas and Montgomery County containing the cities of Akron, Toledo and Dayton respectively. All other counties have conglomerated into the bulky central component in the figure which we will call the main trunk. The demographic plot with these components labeled by their constituent counties is presented in \cref{fig:labelleddemographicplot}.

We note that Summit County appearing in two different connected components seems to simply be a quirk of our choice of hyperparameters, together with the fact that Summit County's financial demographics in years 6-10 of our time period were noticeably different from those of Hamilton and Lucas County in the same time period.

\begin{figure}[tb]
    \centering
    \includegraphics[width=0.75\linewidth]{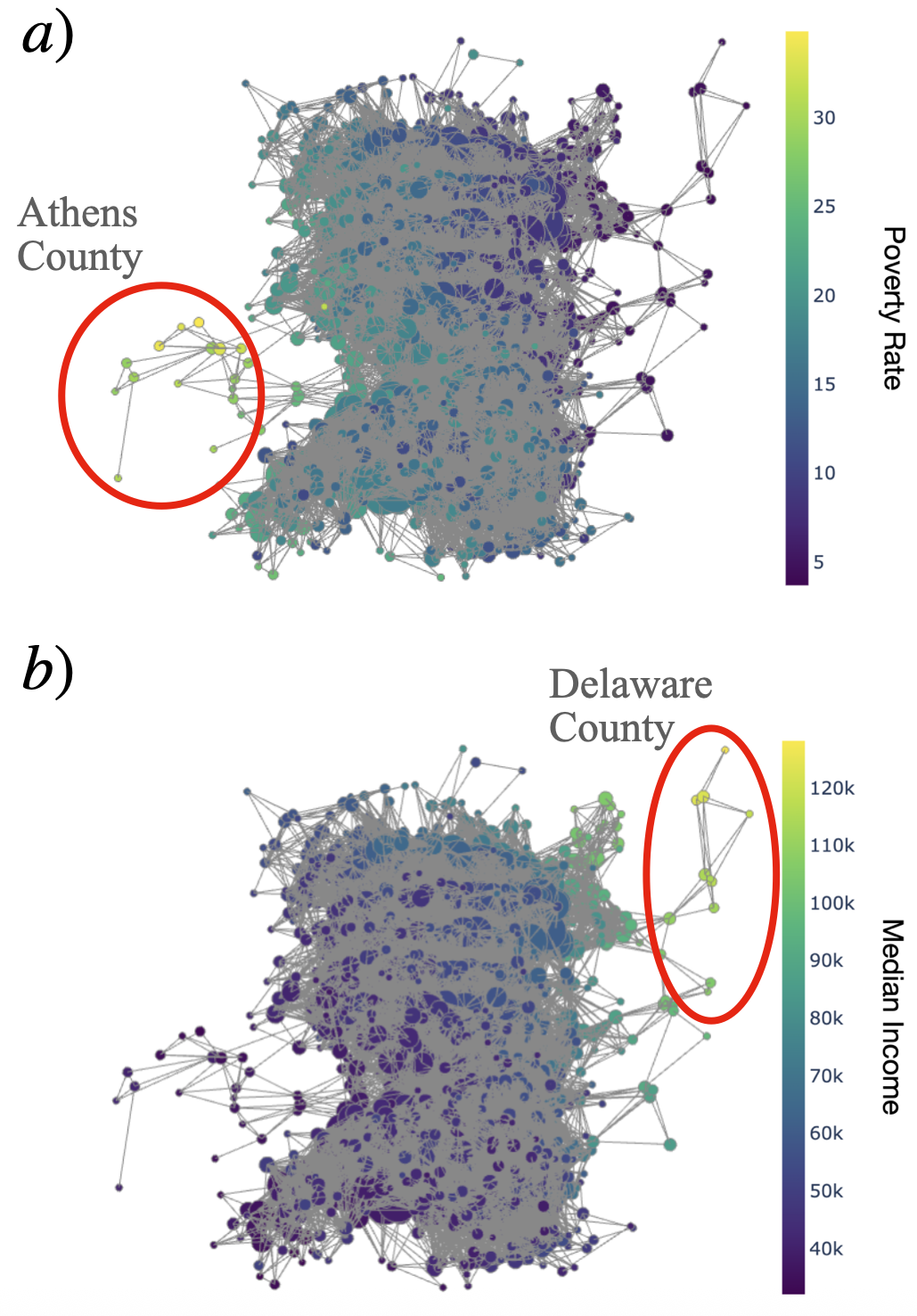}
    \caption{A plot of the visible flares in the Mapper graph formed from the demographic data of the counties of Ohio. a) is colored by poverty rate and b) by median income and shows the formation of negatively correlated gradients within the main trunk.}
    \label{fig:spursindemographicplot}
\end{figure}

When considering the shape of these connected components we can observe that they all seem to flow along a distinct axis of time as evident in \cref{fig:demographic-mapper} a), which should be expected for a linearly increasing variable. By identifying the axis of time in each component we may study how demographics (and eventually death count) change along time within each cluster.
The other demographic features stay relatively constant in the outer connected components, as they represent the data of only a few counties, but seem to create gradients within the main trunk.
This can be seen more clearly by looking at another angle presented in \cref{fig:spursindemographicplot} where we can directly observe a gradient within the main trunk formed by the poverty rate (\cref{fig:spursindemographicplot} a)), which is visually negatively correlated with a gradient formed by median income (\cref{fig:spursindemographicplot} b)).
This figure also explains the other notable geometric feature of these demographic graphs which is the two flares coming off of the main trunk. By node analysis we can see that the flare highlighted in \cref{fig:spursindemographicplot} a) is caused by data from Athens County which had a notably high poverty rate in the first 10 years of our time frame. The flare highlighted in \cref{fig:spursindemographicplot} b) is created by Delaware County which is the wealthiest county in Ohio by a large margin.

The only other demographic feature left to discuss is the unemployment rate. Clusters of nodes with higher unemployment rates seem to form visible ridges on one side of the main trunk of our demographic plot, seen in \cref{fig:demographic-mapper} d). By analyzing the clusters that contribute to these ridges we can see that most of these clusters are composed of counties near the Ohio River, specifically the counties of Adams, Jackson, Meigs, Morgan, Noble, Pike, Scioto and Vinton, as well as some counties in Northern Ohio between Toledo and Cleveland such as Ottawa and Huron.

\subsubsection{Analysis of Overdose-Deaths by Demographic Regions}

When it comes to understanding the connection between this demographic data and the Ohio overdose epidemic there are two approaches for incorporating the overdose data into Mapper graphs.
We can either see how the above Mapper graph is colored according to (normalized) cumulative deaths in these counties over the years, or construct new Mapper graphs formed by including (normalized) cumulative deaths into the filter function.
We have taken both of these approaches in order to contrast the kinds of insights that are gained by both methodologies.

\begin{figure}[tb]
    \centering
    \includegraphics[width=.8\linewidth]{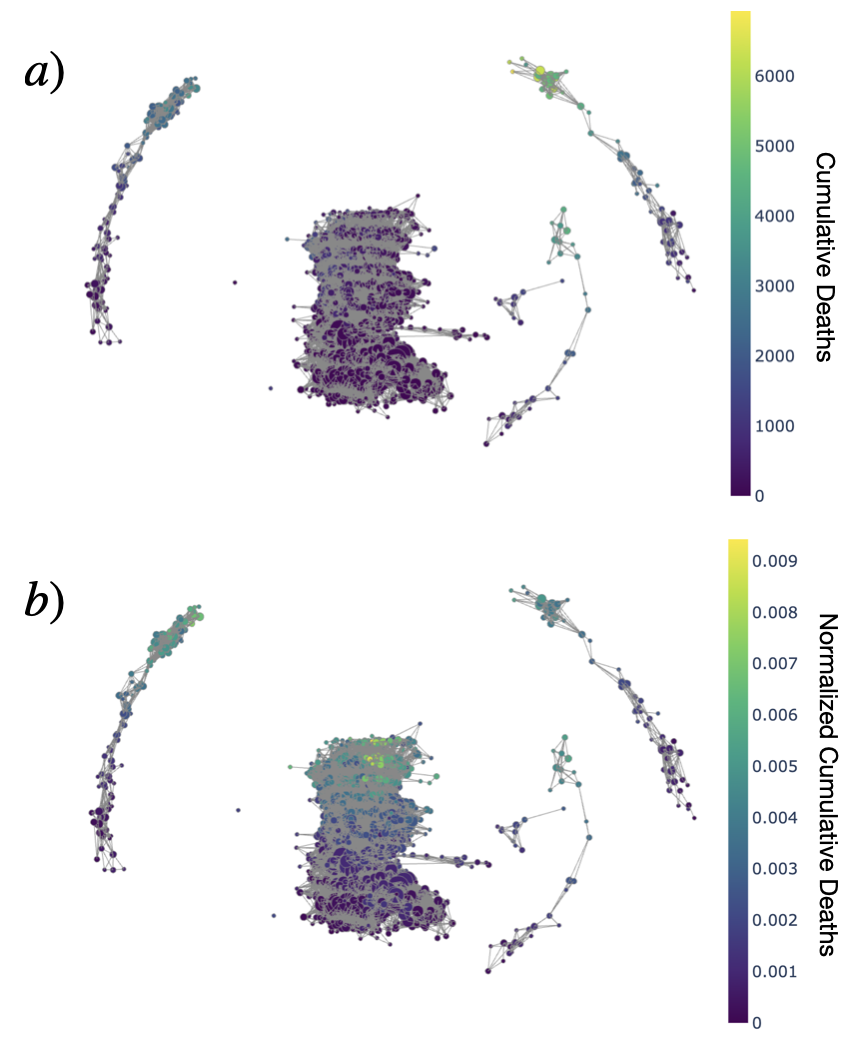}
    \caption{The Mapper graph of demographic data from \cref{fig:demographic-mapper} colored according to the average a) cumulative deaths, b) population-normalized cumulative deaths, of the data points contained in each node.}
    \label{fig:demographic_death_plots}
\end{figure}

We first consider the methodology of coloring our nodes without including death data into our projection filter function. The demographic plot colored according to (normalized) cumulative deaths is presented in \cref{fig:demographic_death_plots}. The immediate takeaway from \cref{fig:demographic_death_plots} a), where we color by cumulative deaths, is that the skinny connected components representing major cities in Ohio experienced higher numbers of cumulative deaths, confirming our earlier analysis of the spatiotemporal plots. When we color by normalized cumulative deaths, \cref{fig:demographic_death_plots} b), we observe that the highest values occur in the main trunk representing the counties of Ohio with smaller populations, though some high values also occur in components containing cities, notably Lucas and Montgomery counties.  

\begin{figure}[tb]
    \centering
    \includegraphics[width=\linewidth]{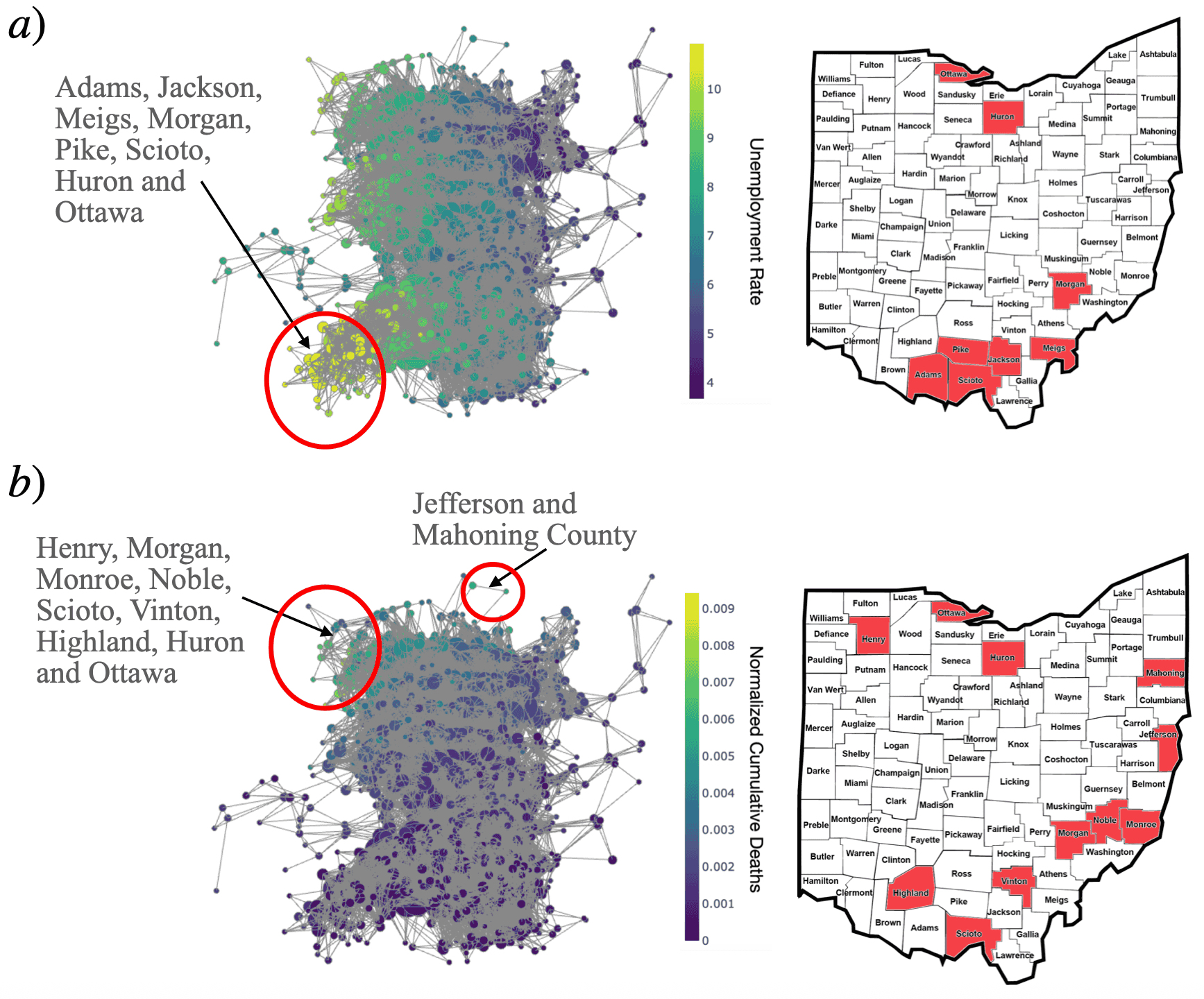}
    \caption{The main trunk of our demographic Mapper graph colored by a) unemployment rate, b) normalized cumulative deaths. Clusters of high values in the respective variables are labeled and the counties whose data contributes to these clusters are highlighted on the adjacent county maps.}
    \label{fig:unemployment_death_comparison}
\end{figure}

By analyzing the clusters where the higher normalized cumulative deaths occur we can see that they are predominantly in areas near the border with Kentucky and West Virginia, specifically Henry, Morgan, Monroe, Noble, Scioto and Vinton counties, along with areas in Northern Ohio outside Toledo like Highland, Huron and Ottawa counties. There is also a small flare from counties along the Pennsylvania border outside the Cleveland area, specifically Jefferson and Mahoning counties as indicated in \cref{fig:unemployment_death_comparison}. All of this aligns with our previous analysis and previous research identifying counties with high overdose death rates \cite{sudors-ohio, rosenblum2020rapidly}.

What is notable about this visualization of normalized cumulative deaths is how it interplays with the unemployment rate as we can see in \cref{fig:unemployment_death_comparison}. 
The cluster which experienced high unemployment at the beginning of our chosen time frame (2009) is located below one of the clusters which experienced high rates of overdose deaths at the end of our chosen time frame (2023) when aligning this main trunk so that the z-axis represents the flow of time. By observing the highlighted county maps in \cref{fig:unemployment_death_comparison} we can see that these clusters are formed from the data of counties that are geographically near each other, with the exception of the Jefferson and Mahoning flares which lie in a different part of the main trunk and are closer to the Pennsylvania border. This all suggests a delayed correlation between these two variables of high unemployment and high overdose deaths which could guide and inform future research into potential predictive correlates.

This pattern is consistent with the broader literature on ``deaths of despair,'' a term popularized by Case and Deaton \cite{case2021deaths} to describe rising mortality from drug overdoses, alcohol-related liver disease, and suicide, particularly among socioeconomically disadvantaged populations. Economic instability, job loss, and declining prospects—especially in regions with persistent poverty and high unemployment—have been strongly associated with increased risk of fatal overdose. Our findings support the hypothesis that economic distress creates long-term vulnerabilities to substance use and overdose, especially in rural and post-industrial communities. This delayed correlation between early economic decline and later spikes in normalized cumulative deaths underscores the importance of proactive economic and public health interventions.

\begin{figure}[tb]
    \centering
    \includegraphics[width=1\linewidth]{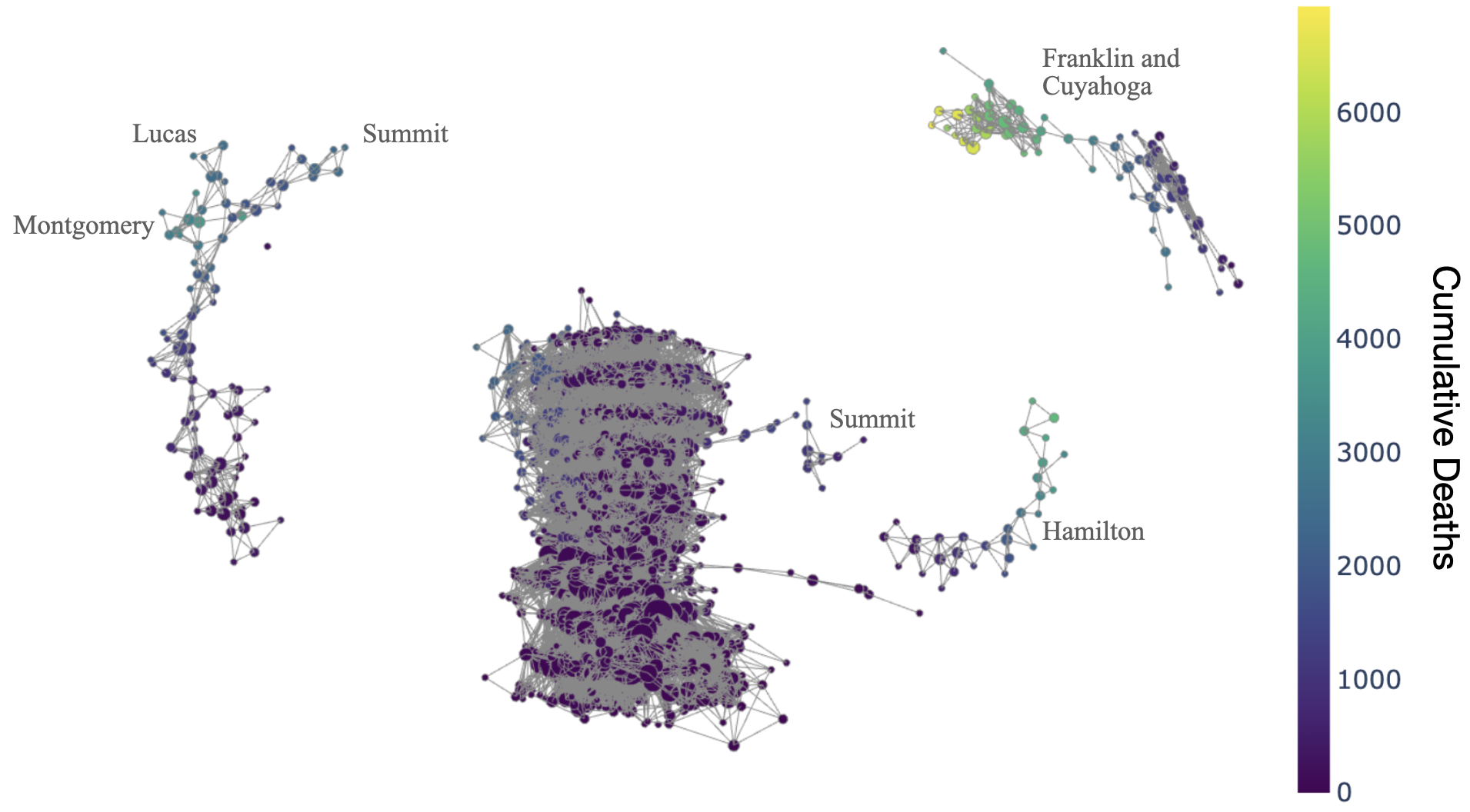}
    \caption{The Mapper graph created by incorporating cumulative death counts into the filter function used for \cref{fig:demographic-mapper} colored by the cumulative death counts.}
    \label{fig:cumulative_death_demographic_projection}
\end{figure}

\begin{figure}[tb]

    \centering
    \includegraphics[width=1\linewidth]{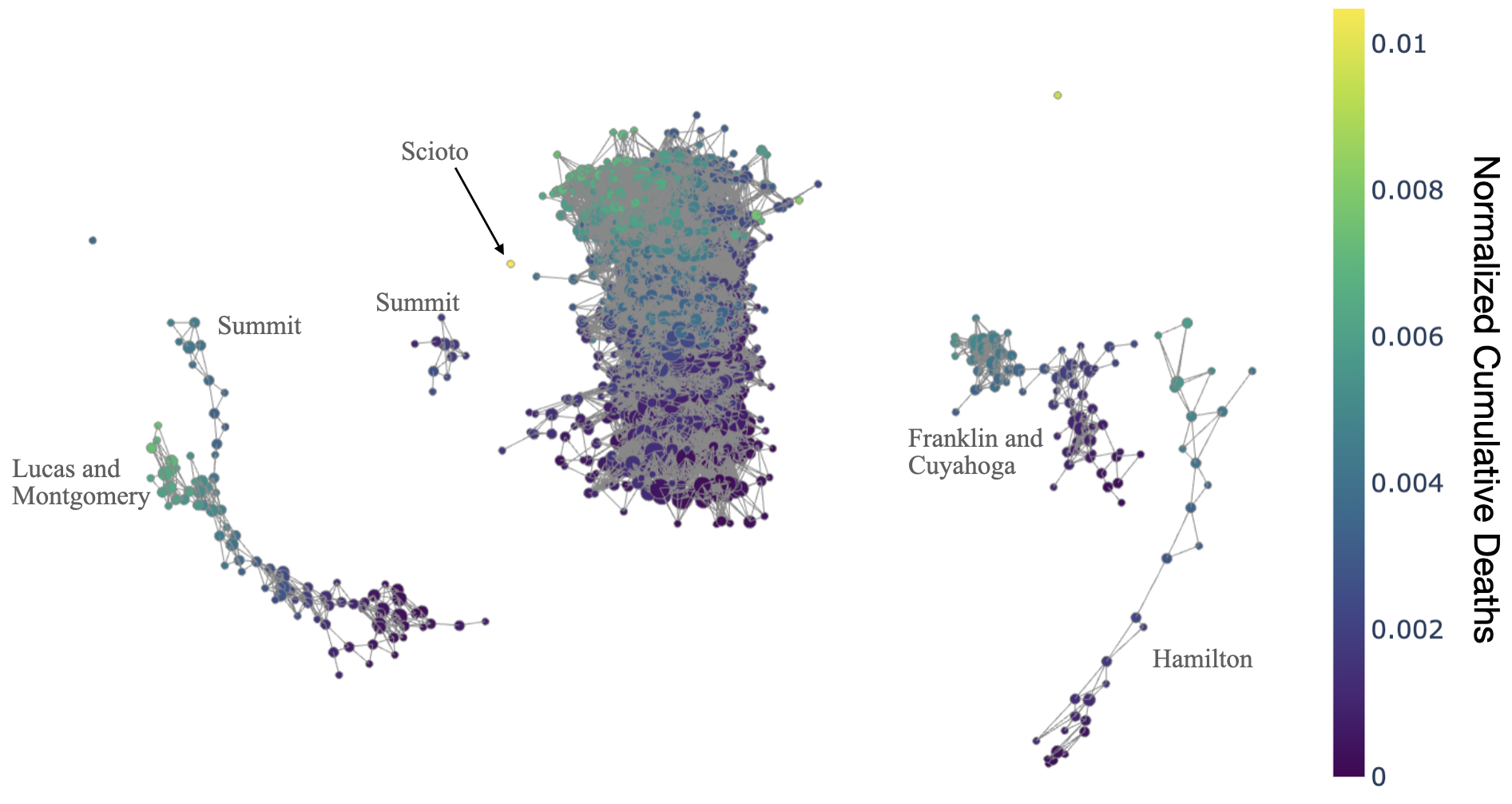}
    \label{fig:demographics+normalizedcumulative}
    \caption{The Mapper graph created by incorporating normalized cumulative death counts into the filter function used for \cref{fig:demographic-mapper} colored by the normalized cumulative death counts.}
    \label{fig:normalized_cumulative_deaths_demographic}
\end{figure}

The plot \cref{fig:demographic_death_plots} b) also highlights the potential for visual occlusion. Some of the nodes with very high values of normalized cumulative deaths are buried inside the main trunk and are hard to see. We can fix this problem by including (normalized) cumulative deaths into our projection filter function, which is the second methodology we considered. The Mapper graphs formed by incorporating cumulative deaths or normalized cumulative deaths into our projection filter function are presented in \cref{fig:cumulative_death_demographic_projection} and \cref{fig:normalized_cumulative_deaths_demographic} respectively.

As a small note, due to sensitivity issues with the Mapper algorithm, in constructing the plot in \cref{fig:normalized_cumulative_deaths_demographic} we had to change our overlap fraction from 45\% to 44\% to clean up the image.

Looking at \cref{fig:cumulative_death_demographic_projection}, very few changes are made to the demographic landscape by incorporating cumulative deaths.
One factor that explains this is that the flares in cumulative deaths came from counties containing major cities in Ohio but as our demographic landscape was already partitioned according to population, our cumulative death information does not cause further delineation.
That is not to say that there is no extra information provided by this plot, as we can see some greater separation between Lucas, Montgomery and Summit counties at the end of our time frame and the emergence of a hole in this connected component towards the start of our time frame. By analyzing nodes we can see that this hole identifies a period in time where Montgomery County experienced a greater increase in deaths than Lucas County before the cumulative death data in both counties came back to having similar values. 
The slight separation at the end of our time frame is also caused by the three counties experiencing different levels of cumulative deaths, where Lucas and Montgomery experienced higher death counts than Summit County. This is notable as it agrees with our earlier investigation of spatiotemporal data where Summit County did not form a visible flare in \cref{fig:CumulativeDeathPlots} while both Montgomery and Lucas County did. 

Similar insights can be gained from \cref{fig:normalized_cumulative_deaths_demographic} where we see an even more significant delineation between Summit County and Lucas and Montgomery County and the appearance of a hole at the same time as the one in \cref{fig:cumulative_death_demographic_projection}.
The major advantage in \cref{fig:normalized_cumulative_deaths_demographic} compared to \cref{fig:demographic_death_plots} b) is that the nodes containing higher death rates are no longer occluded. In fact, the major peak of normalized death count in Scioto County is substantial enough to break off from the main trunk creating its own connected component. This reflects Portsmouth's infamous presence in the drug epidemic as the ``pill mill of America'' \cite{pillmill}.

\section{Conclusions}
\label{section:conclusions}

This paper explored the efficacy of Mapper graphs for studying spatiotemporal data, illustrated by data from the Ohio overdose epidemic.
We confirmed the main findings of \cite{chen2021topological} and conclude that Mapper does a good job finding flares, holes and connected components in spatiotemporal epidemiological data.
Furthermore, we found that we can use prior information of our subject (in our case the geography of Ohio) to inform Mapper hyperparameters in a way that addresses the large number of parameter choices inherent to the algorithm.
These choices allow us to isolate the effect of a variable of interest on the topology and geometry of the Mapper graphs.
We further extended the findings of \cite{chen2021topological} to population-normalized statistics and showed that similar insights can be gained by studying the same kinds of topological and geometric features with this data, yielding new insights as to the impact of the spread on different regions.
This allows us to consider spatiotemporal correlations between densely populated areas and less populated areas. 
In our case, the resulting Mapper graphs visually suggested a delayed correlation between spread in city-centers and surrounding rural areas.

We also introduced a notion of demographic distance between counties, and used Mapper to see how overdose death data interplays with demographic landscapes that shift over time.
We observed that the same features of these Mapper graphs could lead to insights into the dynamics of the Ohio overdose epidemic.
Some of the insights we found were that in terms of raw numbers, the epidemic mostly hurt major urban areas of Ohio, which are the most highly populated. The fatalities started to drastically increase in these areas around the middle of the time frame studied (around 2014/15), coinciding with the introduction of fentanyl to the drug supply.
We also found that areas near to the Ohio River and outside of cities experienced proportionally higher death rates than all other counties and these flares occurred towards the end of the time frame we studied--around 2020.
We were also able to pick out a very strong spike in normalized deaths occurring in Scioto County which contains the town of Portsmouth, a known hotspot in the epidemic.
By studying the demographic profiles we found that areas with low population that experienced higher rates of poverty and unemployment were those hit hardest near the end of our time frame (after normalizing by population).
We also saw a delayed correlation between unemployment rates spiking in an area before spikes in population-normalized overdose deaths.

Overall, we find that Mapper visualizations are effective at quickly highlighting which variables are likely to be worthy of further investigation and which are not by the visual correlations and studying the topological and geographic features of these plots.
This may be a useful tool for researchers wanting to eliminate certain factors from consideration by showing that they bear little relevance to the dependent variable of interest.
Mapper graphs are also compelling visual aids for demonstrating broad trends and features within a spatiotemporal dataset. However, Mapper is first and foremost a visual rather than predictive tool, and the lack of a statistical framework can make it difficult to decide which features are significant.
One main limitation of Mapper is the subjectivity of hyperparameters, which we address in the spatiotemporal case but which remains a problem for reproducibility in the demographic case. 
Developing rigorous sensitivity analyses for these hyperparameters is an important direction for future work with Mapper.
Furthermore while the 3D plots allowed for the nodes to be less cluttered, there are still issues in identifying details due to the visually busy nature of the Mapper embeddings. We encourage further experimentation with this methodology for other spatiotemporal datasets.

As regards the overdose epidemic, we found that Mapper is a powerful exploratory tool for identifying spatial, temporal, and demographic patterns in complex public health data. While they cannot replace statistical significance testing, they effectively spotlight relationships worthy of deeper investigation, such as the delayed correlation between economic stressors and overdose mortality. Given their interpretability and flexibility, we recommend broader application of topological data analysis methods to spatiotemporal public health datasets.

\section*{Supplementary Materials}
\label{sec:supplemental_materials}

The following GitHub contains our data files, interactive HTML Mapper graph visualizations, and the code used to create them:
\url{https://github.com/willeyna/OhioOverdoseMapper}.


\acknowledgments{
The authors are grateful to Ismar Volic for sharing the code from \cite{chen2021topological}, Daniel Rosenblum for sharing data from \cite{rosenblum2020rapidly} regarding important demographics for the Ohio overdose epidemic, and David Kline for sharing code from \cite{kline2021estimating}.

The authors would also like to acknowledge the helpful conversations they had with Gillian Grindstaff and Facundo Memoli to understand more about TDA techniques, and with Andrew Curtis, Jackie Curtis, and Adam Eck regarding spatial models of overdose data. Lastly, the middle author would like to thank his colleagues on the data subcommittee of the US Attorney Heroin/Opiate Task Force for helpful conversations, and for their continuing work to reduce overdose deaths.}

\bibliographystyle{abbrv-doi}
\bibliography{template}

\end{document}